\documentclass[11pt]{article}
\usepackage{geometry}                
\usepackage{amsmath}
\usepackage{graphicx,psfrag,epsf}
\usepackage{enumerate}
\usepackage{natbib}
\usepackage{blindtext, rotating} 
\usepackage{subfig}
\usepackage{amssymb}
\usepackage{booktabs}
\usepackage{epstopdf}
\usepackage[x11names,table]{xcolor}
\usepackage[font=footnotesize,labelfont=bf]{caption}
\title{Time Series Clustering using the Total Variation Distance with Applications in Oceanography}
\author{Pedro C. Alvarez-Esteban$^a$\thanks{Corresponding author. E-mail: pedroc@eio.uva.es. Phone: +34 983 423930}
\hspace{1cm} C. Eu\'an$^b$ \hspace{1cm} J. Ortega$^b$ \\ \\
$^a$ {\small\textit{Dept. de Estad\'{\i}stica e Investigaci\'on Operativa, Universidad de Valladolid.}} \\
\small{\textit{Paseo de Bel\'en, 7. 47005 Valladolid. Spain.}}\\
$^b$ \small{\textit{CIMAT, A.C.  Jalisco, s/n, Mineral de
Valenciana. Guanajuato 36240, Mexico.}}}

\begin{document}
\bibliographystyle{plain}

\maketitle

\begin{abstract}

An algorithm for determining stationary periods for time series of
random sea waves is proposed in this work. This is a problem in
which changes between stationary sea states are usually slow and
segmentation procedures based on change-point detection frequently
give poor results. The method is based on a new procedure for time
series clustering, built on the use of the total variation distance
between normalized spectra as a measure of dissimilarity. The
oscillatory behavior of the series is thus considered the central
characteristic for classification purposes. The proposed algorithm
is compared to several other methods which are also based on
features extracted from the original series and the results show
that its performance is comparable to the best methods available and
in some tests it performs better. This clustering method may be of
independent interest.
\end{abstract}

\noindent%

\noindent\textbf{MSC classes:} 62H30, 62M10, 62M15.

  \maketitle

\section{Introduction}
\label{sec:intro}

An algorithm for determining stationary periods for time series of
random sea waves is proposed in this work. This is a problem in
which changes between stationary sea states are usually slow. The
method is based on a new procedure for time series clustering, built
on the use of the total variation distance between normalized
spectra as a measure of dissimilarity. The oscillatory behavior of
the series is thus considered the central characteristic for
classification purposes.

Random processes have been used to model sea waves since the 1950's,
starting with the work of \citet{Pierson} and \citet{l-h:1}. Models
based on random processes have proved useful, allowing the study of
many wave features \citep[see, e.g.][]{Ochi:ochi}. A class of models
often used to study sea waves in deep waters with standard
conditions are stationary centered Gaussian processes
\citep{Space,Ochi:ochi}. The stationarity hypothesis allows the use
of Fourier spectral analysis to study the wave energy distribution
as a function of frequency. In particular, this spectral analysis is
related to several features of interest, such as the significant
wave height ($H_s$) or the dominant or peak period ($T_p$), that can
be computed from the spectral distribution \citep[see,
e.g.][]{Ochi:ochi}. On the other hand, Gaussian models, beyond being
a good first order approximation, allow obtaining explicit
expressions for the distribution of parameters of interest. However,
both hypotheses, stationarity and Gaussianity, have limitations. It
is clear that in the medium/long-term the sea is not stationary.
Thus, the use of stationary models is limited in time, depending on
the  specific sea conditions at the place of study. In other words,
the sea state at a specific point can be regarded (or modeled) as an
alternating sequence of stationary and transition periods (between
the stationary periods).

The problem of duration of sea states is linked to the detection of
change-points in time series. However, the methods employed to this
effect usually assume that changes in the time series occur
instantaneously or in a very brief period of time, which is not
usually the case for waves, where changes take time to develop. This
problem has been studied  from different points of view.
\citet{OyH1} compared the results of using two methods, detection of
changes by penalized contrasts proposed in \citet{marc1} and
\citet{marc4} and the smoothed localized complex exponentials
(SLEX), introduced by \citet{ORSG}, with unsatisfactory results.
\citet{SyS1} propose a segmentation method for significant wave
height based on determining periods of stability, increase and
decrease using time-series and local regression techniques.
\citet{OyH2} consider a method based on calculating mean values over
moving windows, and using a fixed-width band to determine change
points in the wave-height data. Other studies \citep{SyT1, MyP1,
MAyP1} have focused on the joint distribution of certain wave
parameters, both from the point of view of estimation and from the
point of view of simulation, with the purpose of obtaining duration
distribution parameters through Monte Carlo methods. \citet{Jenkins}
considers the problem from the perspective of estimating the fractal
(Hausdorff) dimension.

As an application of time series clustering, we propose a new method
for determining stationary periods for random waves, that takes into
account the fact that transitions take some time to develop. The
point of view switches from detecting  change points to the
identification of time intervals during which the behavior of the
 time series is stable.  These time series are divided
into 30-minute periods, a time interval which is usually considered
to be long enough for a good estimation of the spectral density and
short enough for stationarity to be a reasonable assumption. The
clustering algorithm is then applied to the set of 30-minute
intervals. If the clusters obtained  are contiguous in time they are
considered to be stationary intervals. The procedure also allows for
the determination of transition intervals between successive
stationary periods. Since one of our goals is to develop a method
for determining stationary time intervals for sea waves that could
be useful in Oceanography, the WAFO toolbox \citep{wafo:wafo} in
Matlab was used for all standard calculations regarding spectral
densities and simulations from parametric spectral families.

In general, clustering is a procedure whereby a set of unlabeled
data is divided into groups so that members of the same group are
similar, while members of distinct groups differ as much as
possible. The problem of clustering when the data points are time
series  has received a lot of attention in recent times.
\citet{Liao05} gives a revision of the field up to 2005 and
\citet{MyV1} present an \texttt{R} package (TSclust) for time series
clustering with a wide variety of alternative procedures. A thorough
revision of the literature in recent years is outside the scope of
this work, but the subject has found applications in diverse fields
such as the identification of similar physicochemical properties of
amino acid sequences \citep{savvides2008clustering}, analysis of
fMRI data \citep{Goutte1999298}, detection of groups of
stocks sharing synchronous time evolutions
 with a view towards portfolio optimization
\citep{basalto2006}, the identification of geographically
homogeneous regions based on similarities in the temporal dynamics
of weather patterns \citep{bc2008} and finding groups of similar
river flow time series for regional classification
\citep{Corduas201173}, to name but a few.

According to \citet{Liao05} there are three approaches to time
series clustering: methods based on the comparison of raw data,
feature-based methods, where the similarity between time series is
gauged through features extracted from the data and methods based on
parameters from models adjusted to the data. Our approach falls in
the second group, and the feature used is the spectral density of
the corresponding time series. The similarity between two time
series is measured by the total variation distance (TV) between
their normalized spectra. This distance is frequently used to
compare probability measures, and requires the normalization of
spectral densities, so that the integral of the normalized density
is equal to one. This is equivalent to normalizing the time series
so that its variance is equal to one. Thus, we focus on differences
in the distribution of the variance as a function of frequency
rather than differences in the total variance. The use of the TV
distance for the analysis of differences in the context of spectral
analysis of random waves was proposed by \citet{AyO1} and also
considered in \citet[][]{EOyA1, EOyA2}.

Once the spectra for the time series have been estimated and
normalized, the TV distance between all pairs are calculated and
used to build a dissimilarity matrix, which is then fed to an
agglomerative hierarchical clustering algorithm. Several linkage
criteria were used and Dunn's index was employed for deciding the
optimal number of clusters.

Many clustering algorithms have been devised for time series and to
compare their performance \citet{Vilar1} proposed a series of tests.
To gauge the efficiency of our algorithm the same tests were used.
Since our interest lies in applications to random wave data, an
additional test using families of spectral densities frequently used
in Oceanography was also carried out. These tests show that, in most
cases, the performance of the proposed algorithm compares with the
best available, and in some cases it outperforms the rest.

The rest of this article is organized as follows: Section 2
introduces the TV distance, which will be used as the similarity
measure between normalized spectral densities for the time series.
Section 3 describes the clustering algorithm based on the TV
distance. Section 4 reports results from a simulation study based
partly on \citet{Vilar1} to compare
the clustering algorithm with other methods. In Section 5 two types
of applications are considered, first, using simulated data that
includes transition periods the performance of the algorithm is
assessed, and second, an application to real wave data is discussed
in detail. The paper ends with conclusions about the results
obtained.

\section{Total Variation Distance}
The total variation (TV) distance is one of the most widely used
metrics between probability measures. Although it can be defined in
general probability spaces, we will focus on the real line, $\mathbb
R$. Given $P$ and $Q$, two probability measures in $\mathbb R$, the
total variation distance between them is defined as
\begin{equation} d_{TV}(P,Q) = \sup \{|P(A) - Q(A)|: A\in\mathcal
B\} \label{ecu2-1}
\end{equation}
where $\mathcal B$ is the class of the Borel sets
on the real line.

One important property of the TV distance is that it is bounded
between 0 and 1, being 1 the largest possible distance between two
given probabilities. This property can be easily deduced from the
definition. Obviously it is positive, and taking into account that
for every Borel set $A$, $0\leq P(A), Q(A) \leq 1$  then, $0\leq
|P(A) - Q(A)| \leq 1$ and the inequalities remain valid if we take
the supremum over the sets in $\mathcal B$. A value of 1 for the
distance can be attained if $P$ and $Q$ have disjoint supports.

This property is very useful in order to interpret distance values
between two probabilities: values close to 1 mean that the two
probabilities are quite different while distance values close to 0
mean that these probabilities are very similar, almost equal. A
statistical test to contrast the null hypothesis that the TV
distance between two probabilities is less or equal to a given
threshold has been recently developed in  \citet{Pedro1}.

If $P$ and $Q$ have density functions (typically with respect to the
Lebesgue measure $\mu$), $f$ and $g$, the TV distance between them
can be computed \citep[see, e.g.,][]{Massart} using the
following expression:
\begin{equation}
d_{TV}(P,Q)  = 1 - \int_{-\infty}^\infty \min(f,g)\, d\mu =
\frac{1}{2} \int |f-g|\, d\mu
\end{equation}
and the supremum in (\ref{ecu2-1}) is attained with the set
$A=\{f>g\}$.

\begin{figure}
\centering
 \includegraphics[height=5cm]{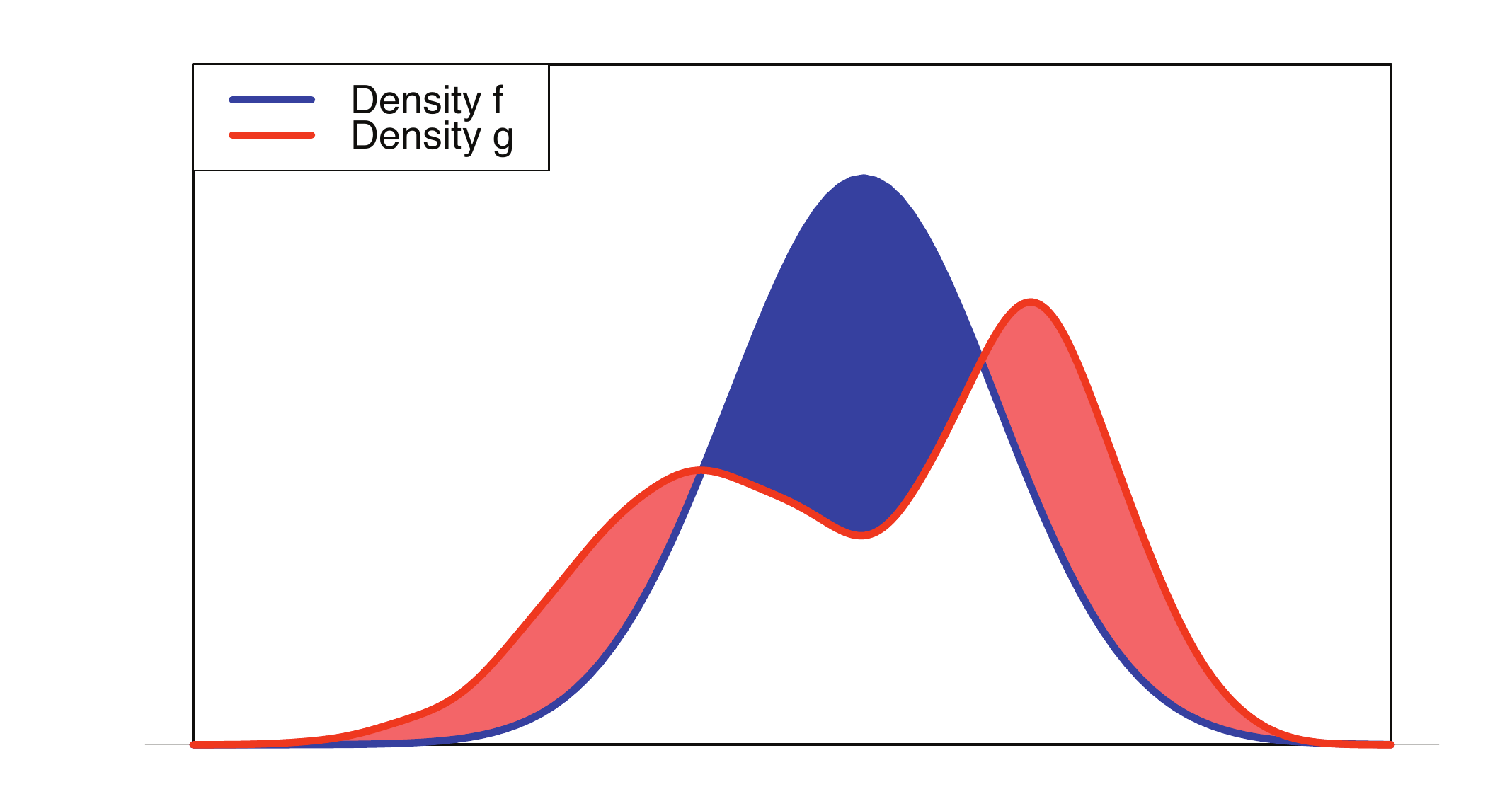}
 \caption{The TV distance measures the similarity between the two densities.
 The blue (orange) area is the value of the TV distance.}
 \label{F_VT1}
\end{figure}

This equation helps to graphically interpret the TV distance. If two
densities $f$ and $g$, have TV distance equal to $1-\alpha$ this
means that they share a common area of size $\alpha$. Thus, the more
they overlap the closer they are. Figure \ref{F_VT1} illustrates the
case with two density functions and shows how to compute the TV
distance. In this figure, the area of the orange region represents
the TV distance, and is equal to the blue area, since the area under
both densities is 1. Both colored regions represent the non-common
part of the density functions, while the white area under the curves
is the common part.

An alternative to the TV distance that is frequently considered is
the Kullback-Leibler divergence between $P$ and $Q$, which is
defined as
\[
K(P,Q) = \int f \log \frac{f}{g}\, d\mu.
\]
$K(P,Q)$ is not a true distance, but it has similar properties and
is useful in many contexts. $K$ is always non-negative and takes the
value 0 if and only if $P=Q$. However, if the set $\{f>0, g=0\}$ has
strictly positive Lebesgue measure, which happens when $P$ is not
dominated by $Q$, then $K(P,Q)=\infty$, while the TV variation
distance between this measures is still bounded by 1. This
situation, in which the K-L divergence is not useful, is feasible in
the type of applications we will be considering. The TV distance is
related to the K-L divergence through the following inequality,
known as Pinsker's inequality
\[
\frac{1}{2} d_{TV}^2(P,Q) \leq K(P,Q).
\]

A discussion of the use of the K-L divergence in discrimination and
cluster analysis of time series using the spectral density can be
found in \citet{SS1}

\section{Clustering using the spectral densities.}\label{sec3}
Our approach to stationary time series clustering is based on the
spectral density as a feature that sums up the oscillatory behavior
of the series around its mean value. In a physical context, e.g.
when considering measurements of sea surface height at a fixed
point, the spectrum of the time series is interpreted as the
distribution of the energy as a function of frequency. The integral
of the spectral density is (proportional to) the total energy
present, and is, of course, the variance of the series. Thus a
normalization of the spectral density corresponds to a consideration
of the frequency distribution of the energy, disregarding the total
energy present. Spectral densities that are similar after
normalization correspond to times series that have similar
oscillatory behavior around their mean values, but may differ in
variance.

Several clustering methods based on spectral densities have been
proposed  in the literature. \citet{S-K} consider periodograms
normalized by dividing by the largest value and use the Euclidean
distance between them to build a dissimilarity matrix, which is then
fed to hierarchical clustering algorithm. \citet{Shumway1} considers
time-varying spectra within the framework of local stationarity, and
uses the Kullback-Leibler discrimination measure, integrated over
both frequency and time, to discriminate between seismic data coming
from earthquakes and explosions. \citet{Caiado1} propose metrics
based on the normalized periodogram for distinguishing stationary
from non-stationary time series. \citet{savvides2008clustering} use
dissimilarity measures based on the cepstral coefficients, which are
the coefficients in the Fourier expansion of the log spectrum.
\citet{Maharaj1} also use cepstral coefficients for a clustering
algorithm based on fuzzy logic.

 \citet{KST98}  propose a general spectral disparity
measure given by
\begin{equation}\label{KST}
d_W(X,Y)  = \frac{1}{4\pi} \int_{-\pi}^\pi W\Big(
\frac{f_X(\lambda)}{f_Y(\lambda)}\Big) d\lambda
\end{equation}
where $f_X$ and $f_Y$ denote the spectral densities of the series
$X$ and $Y$ respectively, and $W$ is a function that must satisfy
certain regularity conditions in order to ensure that $d_W$ has the
properties of a distance, except for the triangle inequality, and is
thus a quasi-distance. With appropriate choices for $W$ (see
\citet{KST98} for details) one can obtain the limiting spectral
approximations to the Kullback-Leibler divergence and the Chernoff
information in the time domain. They use these disparity measures
with a hierarchical clustering algorithm and consider an application
in seismology.

Our approach considers the normalized spectra of the time series as
the feature of interest for clustering, and the TV distance between
them is used to measure the similarity between two time series. The
spectral densities were estimated using the inverse Fourier
transform of the ACF, smoothed using a Parzen window with a
bandwidth of length 100, with the toolbox WAFO in Matlab.

To choose the bandwidth, a series of test were performed. Using
spectra from several parametric families of frequent use in
Oceanography \citep[see][]{JONSWAP,Torset1,Torset2,Ochi:ochi} a Gaussian process was
generated having a given spectral density, using WAFO. The data
generated had a sampling frequency of 1.28 Hz. and corresponded to a
30-minute period, parameters that agree with the usual conditions
for measurements in sea buoys. The empirical spectral density
function was estimated using a range of bandwidths, and the TV
distance between the original and the estimated spectral densities
was calculated. This procedure was repeated 1000 times. The results
showed that a bandwidth value around 100 was optimal.

The proposed clustering procedure is as follows:
\begin{itemize}
\item For each time series the spectral density is estimated and normalized.

\item The total variation distance between the normalized spectral
densities are calculated and used to build a dissimilarity matrix.

\item This dissimilarity matrix is fed to an agglomerative
hierarchical clustering algorithm. We considered two different
linkage criteria: complete and average, and used the function
\textit{agnes} in R \citep{R}.

\item To choose the number of clusters when no external indication
was available, Dunn's index  was used. See section \ref{sec5-2} for
details.

\end{itemize}

\section{Simulations}\label{sec:simulations}
\citet{Vilar1} proposed two simulation tests to compare the
performance of several clustering algorithms. These tests were
reproduced here to compare the performance of the algorithm proposed
in this article, with the best algorithms available. A third test
based on simulated waves was added. In their study, P\'ertega and
Vilar considered several dissimilarity criteria. For our purpose, we
only included those that were not model-based and had the best
results, plus the distance based on the cepstral coefficients (the
Fourier coefficients of the expansion of the logarithm of the
estimated periodogram). The dissimilarity criteria in the time
domain included were:
\begin{itemize}
\item The distance between the estimated autocorrelation functions with uniform weights:
$ d_{ACFU}(X,Y) = \Big( \sum_i\big(
\hat\rho_{i,X}-\hat\rho_{i,Y}\big)^2\Big)^{1/2} $.

\item The distance
between the estimated autocorrelation functions with decaying
geometric weights:
$
d_{ACFG}(X,Y) = \Big( \sum_i p(1-p)^i\big(
\hat\rho_{i,X}-\hat\rho_{i,Y}\big)^2\Big)^{1/2}
$
with $0<p<1$.
\end{itemize}

Let $ I_X(\lambda_k) = T^{-1} \Big|\sum_{t=1}^T X_t e^{-i\lambda_k
t}\Big|^2 $ be the periodogram for time series $X$, at frequencies
$\lambda_k=2\pi k/T, \ k=1, \dots, n$ with $n=[(T-1)/2]$, and $NI_X$
be the normalized periodogram, i.e. $NI_X(\lambda_k) =
I_X(\lambda_k)/\hat\gamma_0^X $, with $\hat\gamma_0^X$ the sample
variance of time series $X$. The dissimilarity criteria in the
frequency domain considered were:
\begin{itemize}
\item The Euclidean distance between the estimated periodogram ordinates:
$ d_{P}(X,Y) = \frac{1}{n}\Big( \sum_k \big(
I_{X}(\lambda_k)-I_{Y}(\lambda_k)\big)^2\Big)^{1/2}. $

\item The Euclidean distance between the normalized estimated periodogram ordinates:
$ d_{NP}(X,Y) = \frac{1}{n}\Big( \sum_k \big( NI_{X}(\lambda_k)-
NI_{Y}(\lambda_k)\big)^2\Big)^{1/2}.
 $

\item The Euclidean distance between the logarithms of the estimated
periodograms
$ d_{LP}(X,Y) = \frac{1}{n}\Big( \sum_k \big( \log I_{X}(\lambda_k)-
\log I_{Y}(\lambda_k)\big)^2\Big)^{1/2}. $

\item The Euclidean distance between the logarithms of the normalized estimated
periodograms
$ d_{LNP}(X,Y) = \frac{1}{n}\Big( \sum_k \big( \log
NI_{X}(\lambda_k)- \log NI_{Y}(\lambda_k)\big)^2\Big)^{1/2}. $

\item The square of the Euclidean distance  between the cepstral coefficients
$
d_{CEP}(X,Y) =  \sum_k^p \big( \theta_k^X-\theta_k^Y\big)^2
$
where, $\theta_0=\int_0^1 \log I(\lambda) \mbox{d}\lambda$ and
$\theta_k=\int_0^1 \log I(\lambda) \cos(2 \pi k \lambda )
\mbox{d}\lambda$.

\end{itemize}

These dissimilarity measures were compared with the TV distance and
the $L^1$ distance of the log of the normalized spectra, which is
given by
$$
  d_{L^1}(f_1,f_2)=\frac{1}{2}\int |\log(f_1(\omega))-\log(f_2(\omega))|\mbox{d}\omega.
$$
Three experiments were carried out, the first two were those
proposed by P\'ertega and Vilar, and the third used
simulated wave data. The steps for each experiment were:
\begin{enumerate}
 \item Generate a group of time series of length $T$ that have some special characteristic,
 in order to have well-defined groups.

 \item Calculate the dissimilarity matrix for each of the different measures. Here we fix
 some of the parameters as follows: For the ACFG and ACFU distances the maximum lag is $25$
 and for the geometric weights we take $p=0.05$; for the CEP measure we take
 $p=128$ and the spectra were estimated as described in section
 \ref{sec3}.

 \item The dissimilarity matrix is then used in a hierarchical clustering algorithm with the complete link.

 \item The final groups are formed from the dendrogram by fixing the number $k$ of groups.

 \item In order to evaluate the rate of success in the $m$-th iteration, the following index was used.
 Let $\{G_1,\ldots,G_g\}$ and $\{C_1,\ldots,C_k\}$, be the set of
 the $g$ real groups and a $k$-cluster solution, respectively. Then,
 $$
 \mbox{Sim}(C,G)=\frac{1}{g}\sum_{i=1}^{g} \max_{1\leq j\leq k} \mbox{Sim}(C_j,G_i),
 $$
 where
 $$
 \mbox{Sim}(C_j,G_i)=\frac{2|C_j \cap G_i|}{|C_j|+|G_i|}.
 $$
 This was calculated for each trial and the average value is reported in the tables.
\end{enumerate}
\noindent\textbf{Experiment 1.} In this experiment, a series of
ARIMA models are considered. In each iteration, we simulate one
realization of size $T=200$, from each of the following 12 ARIMA
models proposed by \citet{Caiado1}, six of which are stationary and
six non-stationary.
$$ \begin{array}{llll}
 \mbox{a) AR(1)}  & \phi_1=0.9 &\quad  \mbox{g) ARIMA(1,1,0)}  & \phi_1=-0.1 \\
 \mbox{b) AR(2)}  & \phi_1=0.95, \phi_2=-0.1 &\quad  \mbox{h) ARIMA(0,1,0)}  &  \\
 \mbox{c) ARMA(1,1)} & \phi_1=0.95, \theta_1=0.1 &\quad  \mbox{i) ARIMA(0,1,1)} &\theta_1=0.1\\
 \mbox{d) ARMA(1,1)} & \phi_1=-0.1, \theta_1=-0.95 &\quad  \mbox{j) ARIMA(0,1,1)} &\theta_1=-0.1\\
 \mbox{e) MA(1)}   & \theta_1=-0.9 &\quad  \mbox{k) ARIMA(1,1,1)}   &  \phi_1=0.1, \theta_1=-0.1\\
 \mbox{f) MA(2)}   & \theta_1=-0.95, \theta_2=-0.1 &\quad  \mbox{l) ARIMA(1,1,1)}   & \phi_1=0.05, \theta_1=-0.05\\
 \end{array}
$$

\begin{figure}
\centering
 \includegraphics[width=7cm, height=5cm]{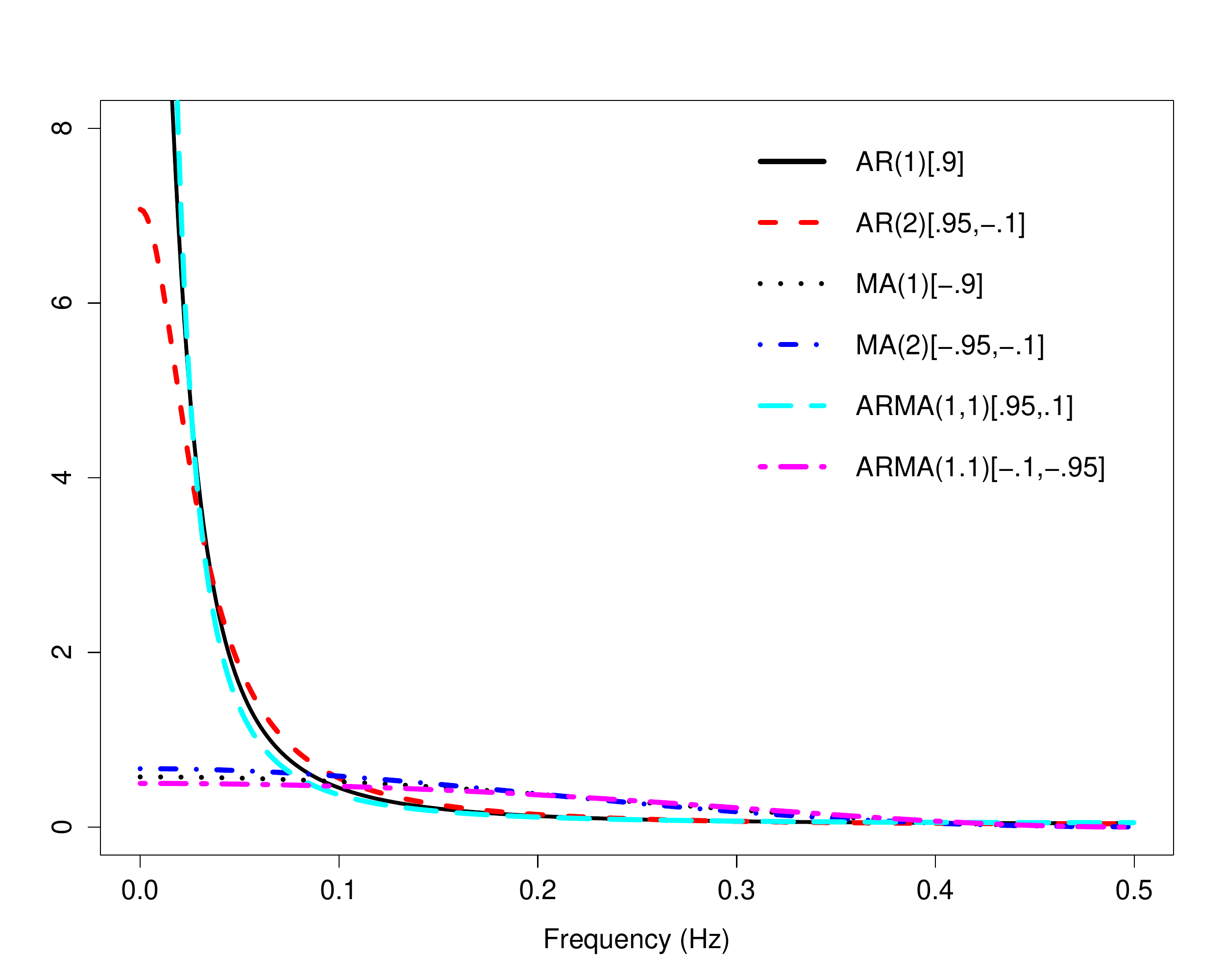}
 \includegraphics[width=7cm, height=5cm]{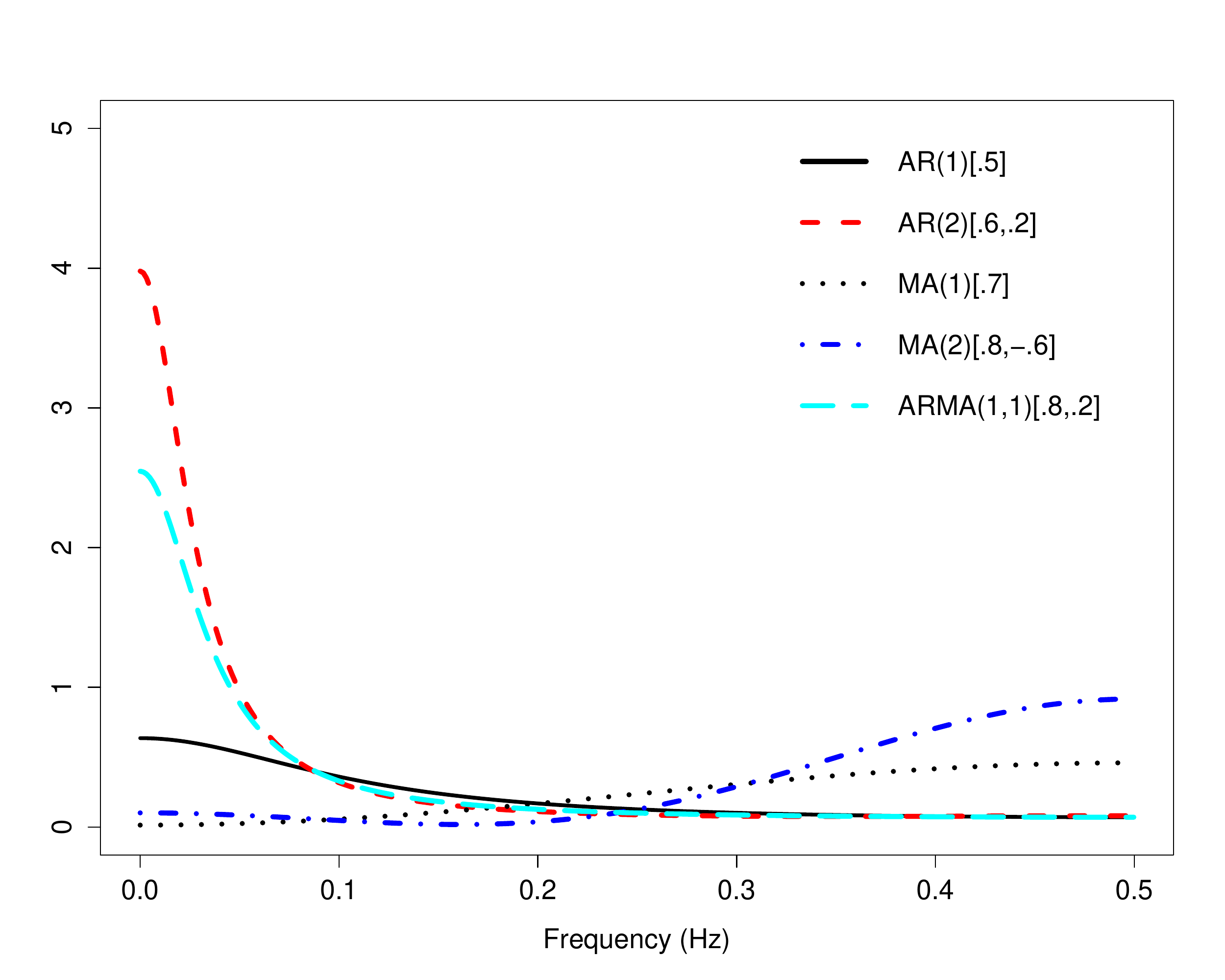}

 \caption{Spectra involved in experiment 1, stationary group (left) and in experiment 2 (right).}
 \label{S_Exp1}
\end{figure}

It is expected that clustering will divide the 12 series into two
groups: stationary and non-stationary. Figure \ref{S_Exp1} (left)
presents the spectral densities  for the stationary processes. The
figure shows that these spectra are not similar and so for the
spectral methods we do not expect to get good results. Table
\ref{tabla:Exp1} shows the  rate of success, the ACFU gets the best
results. A closer examination of the results, not included in the
table, showed that when the spectra are not similar the TV distance
works equally well.

\renewcommand{\arraystretch}{0.8}
\begin{table}
\centering \footnotesize
\begin{tabular}{ l| l l l l l l l l l}
\multicolumn{10}{l}{$T=200$}\\
\hline \hline
\textbf{N} &ACFG    &ACFU   &P  &NP &LP &LNP    &CEP    &TV &L$^1$ \\
\hline
300    &0.859   &\underline{0.873}  &0.667  &0.863  &0.750  &0.751  &0.750  &0.750  &0.756\\
500    &0.859   &\underline{0.876}  &0.671  &0.866  &0.750  &0.750  &0.750  &0.751  &0.756\\
1000   &0.861   &\underline{0.878}  &0.674  &0.870  &0.750  &0.750  &0.750  &0.751  &0.756\\
\hline
\end{tabular}
\caption{Results for Experiment 1. $T$ is the length of the series,
$N$ is the number of replications. The best result for each value of
$N$ is underlined.} \label{tabla:Exp1}
\end{table}

\noindent\textbf{Experiment 2.}  In this case, 5 different ARMA
models were considered, but with a different objective. For each
model four series are generated, and the clustering algorithm is
then applied to the 20 samples, to see if they are able to recover
the original groups. The number of groups in the clustering
algorithm is set to 4 and 5. The 5 ARMA models have the following
parameters: (a) AR(1): $\phi_1=0.5$, (b) MA(1): $\theta_1=0.7 $, (c)
AR(2): $\phi_1=0.6, \phi_2=0.2$, (d) MA(2): $\theta_1=0.8,
\theta_2=-0.6$, (e) ARMA(1,1): $\phi_1=0.8, \theta_1=0.2$.

Figure \ref{S_Exp1} (right) shows the spectral densities for the
five models.  As can be seen, the spectra for the MA models are very
similar so it may be difficult to distinguish them. In this case we
take series of length $T=200, 500$ and $1000$. The results are shown
in Table \ref{tabla:Exp2}.

\renewcommand{\arraystretch}{0.8}
\begin{table}
\centering \footnotesize
\begin{tabular}{ l| l|l l l l l l l l l}
\multicolumn{11}{l}{$T=200$}\\
\hline \hline
$k$&    N&  ACFG&   ACFU&   P&  NP& LP& LNP&    CEP&   TV&  L$^1$\\
\hline
4&  100&    0.530&  0.473&  0.432&  0.469&  \underline{0.722}&  0.701&  0.613&  0.599&  0.703\\
4&  500&    0.536&  0.484&  0.440&  0.480&  \underline{0.718}&  0.695&  0.612&  0.599&  0.699\\
5&  100&    0.660&  0.620&  0.515&  0.610&  \underline{0.939}&  0.736&  0.719&  0.742&  0.925\\
5&  500&    0.663&  0.620&  0.518&  0.611&  \underline{0.928}&  0.739&  0.711&  0.739&  0.922\\
\hline  \multicolumn{11}{l}{\ } \\
\multicolumn{11}{l}{$T=500$}\\
\hline \hline
$k$&    N&  ACFG&   ACFU&   P&  NP& LP& LNP&    CEP&   TV&  L$^1$\\
\hline
4&  100&    0.592&  0.568&  0.490&  0.561&  \underline{0.733}&  0.732&  0.711&  0.664&  0.730\\
4&  500&    0.585&  0.561&  0.492&  0.558&  \underline{0.733}&  0.732&  0.708&  0.667&  0.731\\
5&  100&    0.745&  0.683&  0.561&  0.687&  \underline{0.998}&  0.798&  0.820&  0.852&  0.995\\
5&  500&    0.741&  0.685&  0.566&  0.683&  \underline{0.999}&  0.798&  0.817&  0.846&  0.996\\
\hline \multicolumn{11}{l}{\ } \\
\multicolumn{11}{l}{$T=1000$}\\
\hline \hline
$k$&    N&  ACFG&   ACFU&   P&  NP& LP& LNP&    CEP&  TV&   L$^1$\\
\hline
4&  100&    0.614&  0.583&  0.537&  0.582&  \underline{0.733}&  0.733&  0.730&  0.708&  \underline{0.733}\\
4&  500&    0.615&  0.586&  0.536&  0.587&  \underline{0.733}&  0.733&  0.730&  0.714&  \underline{0.733}\\
5&  100&    0.806&  0.728&  0.600&  0.714&  \underline{1.000}&  0.800&  0.873&  0.907&  \underline{1.000}\\
5&  500&    0.805&  0.737&  0.604&  0.719&  \underline{1.000}&  0.800&  0.874&  0.914&  \underline{1.000}\\
\hline
\end{tabular}
\caption{Results for Experiment 2. $T$ is the length of the series,
$k$ the number of clusters and $N$ the number of replications. The
best result for each value of $T$ is underlined.} \label{tabla:Exp2}
\end{table}

The LP distance works better for small or moderate-length series,
however as $T$ increases the difference with the $L^1$ distance
diminishes, and for $T=1000$ the results are equally good. If we
only compare the spectral distances that do not use the logarithm,
the TV distance is better, with a success rate that is between
$10\%$ and  $20\%$ higher than the rest, including the ACF
distances.

The methods that involved the logarithm of the spectra did not
perform well when the original spectral densities were very close
and the shape was similar. In order to explore this in more detail,
we performed a third simulation experiment, based on parametric
spectra that are frequently used in Oceanography.

\medskip
\noindent\textbf{Experiment 3.} The last experiment is based on two
different JONSWAP (Joint North-Sea Wave Project) spectra. This is a
parametric family of spectral densities which is frequently used in
Oceanography, and is given by the formula
 $$
 S(\omega) = \frac{g^2}{\omega^5}\exp(-5\omega_p^4/4\omega^4)
 \gamma^{\exp(-(\omega-\omega_p)^2/2\omega_p^2s^2)}
 $$
where  $g$ is the acceleration of gravity, $s = 0.07$ if $\omega\leq
\omega_p$ and $s=0.09$ otherwise; $\omega_p=\pi/T_p$ and $\gamma =
\exp(3.484(1-0.1975 (0.036-0.0056 T_p/\sqrt{H_s}) T_p^4/(H_s^2)))$.
The parameters for the model are the significant wave height $H_s$,
which is defined as 4 times the standard deviation of the series,
and the spectral peak period $T_p$, which is the period
corresponding to the modal frequency of the spectrum. This spectral
family was empirically developed after analysis of data collected
during the Joint North Sea Wave Observation Project, JONSWAP,
\citep{JONSWAP}. It is a reasonable model for wind-generated seas
when $3.6\sqrt{H_s} \leq T_p \leq 5\sqrt{H_s}$.

 \begin{figure}
\centering
 \includegraphics[width=7cm, height=5cm]{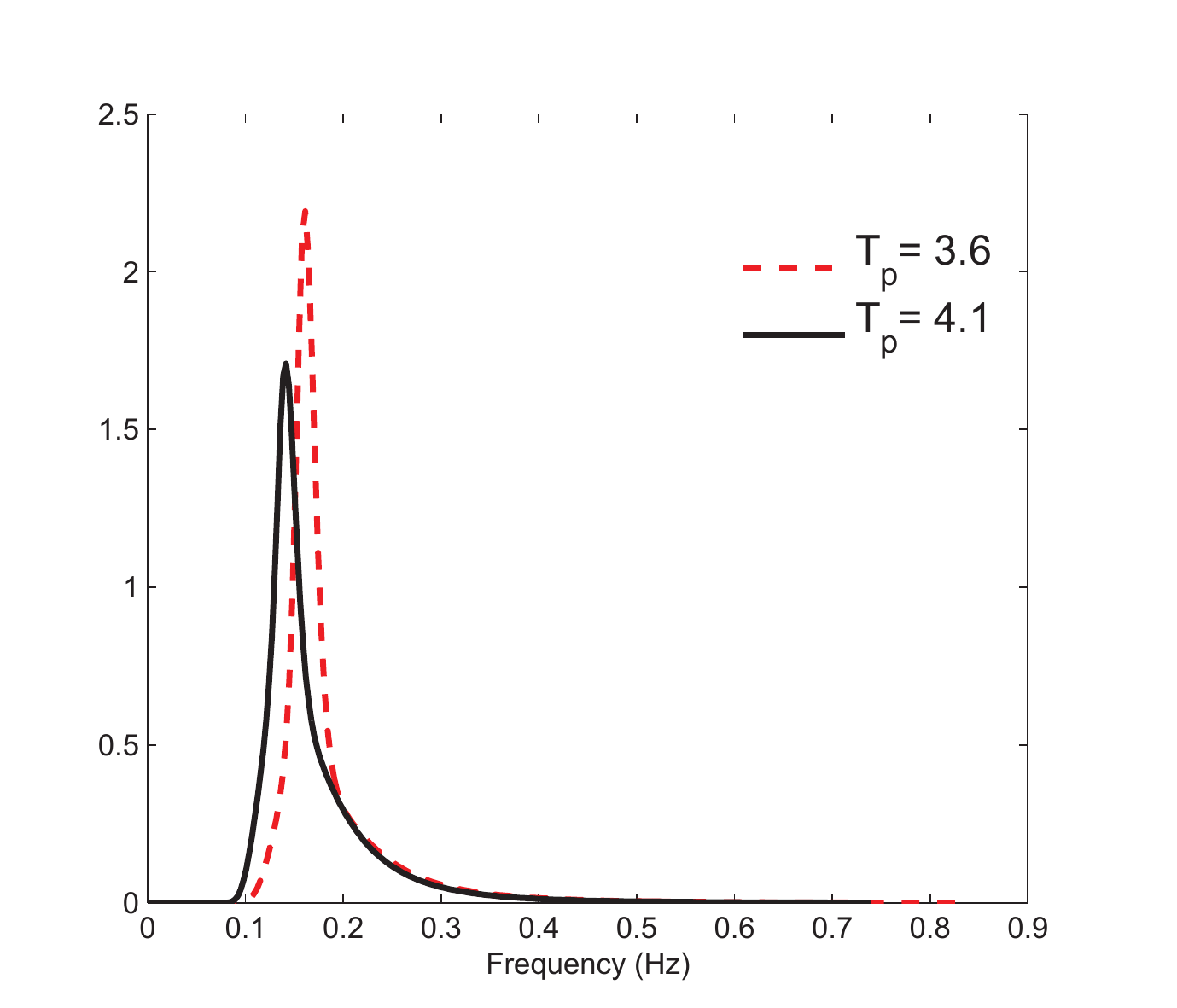}
 \includegraphics[width=7cm, height=5cm]{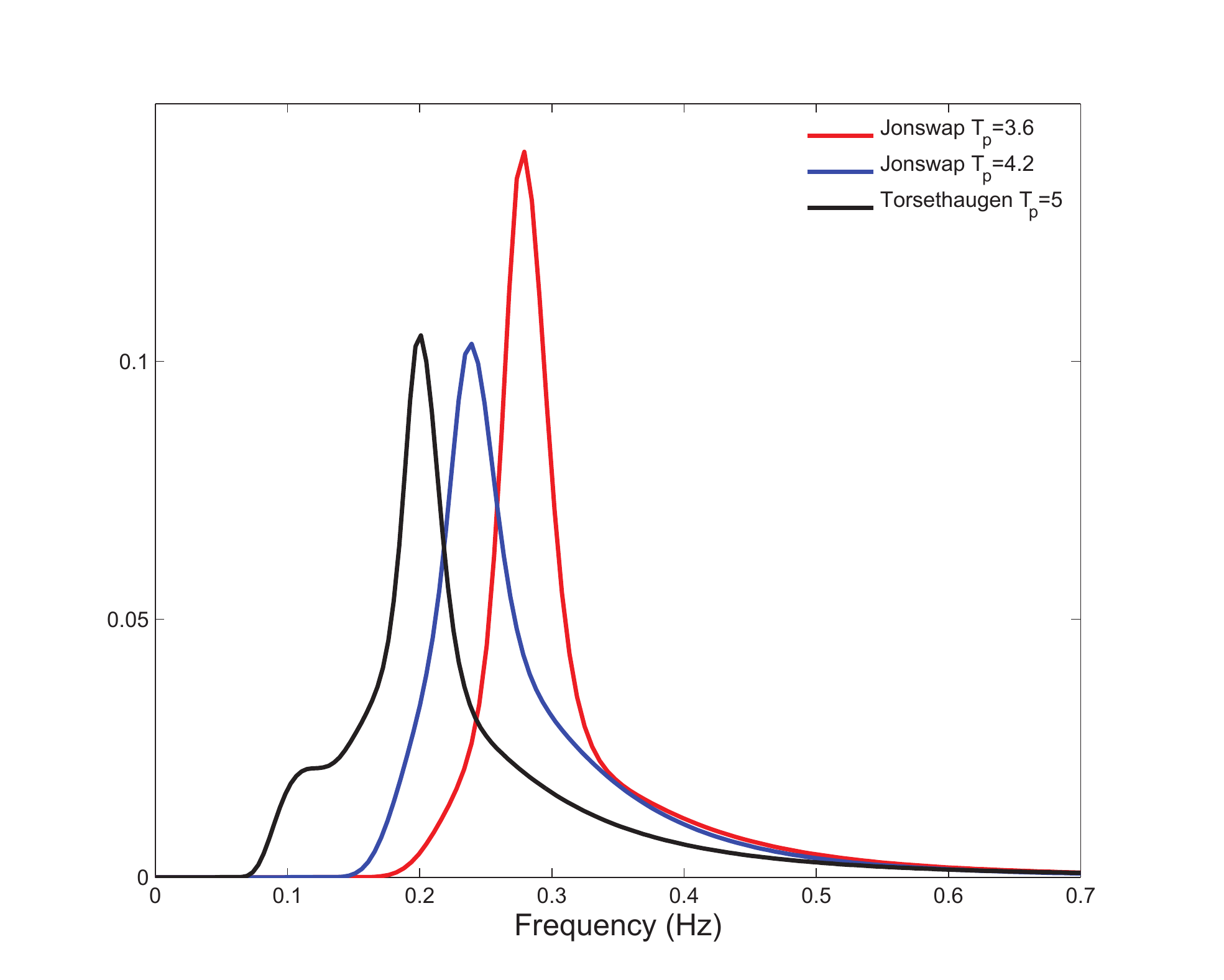}
 \caption{The JONSWAP spectra involved in experiment 3 (left) and
 the spectra involved in the simulations based on transitions (right).}
 \label{S_Exp3}
\end{figure}

The spectra considered both have significant wave height $H_s$ equal
to $3$, the first has a peak period $T_p$ of $3.6\sqrt{H_s}$ while
for the second $T_p= 4.1\sqrt{H_s}$. Figure \ref{S_Exp3} (left)
exhibits the JONSWAP spectra, showing that the curves are close to
each other. This had the purpose of testing the performance of
methods involving logarithms (LP and LNP), which had the best
results in experiment 2, in a different scenario. For data coming
from similar spectra, sampling variability in the estimation of the
spectral densities may be enhanced by the logarithm and this may
have the effect of making more difficult the correct identification
of the two groups.

Four more dissimilarity measures were included in this experiment.
These measures were considered by \citet{Vilar1} in their simulation
study, but did not perform well in the previous experiments. Two of
these measures come from (\ref{KST}) using $\widetilde W(x) = W(x) +
W(1/x)$ with
$$
W(x) = \log(\alpha x+(1-\alpha)) -\alpha\log x)\quad \text{with }
0<\alpha<1.
$$
\citet{Vilar2} studied the asymptotic properties of this estimator
when nonparametric estimators of the spectral densities obtained
using local regression replace the spectral densities in
(\ref{KST}). The two versions of this estimator considered in this
study were $d_{W(DLS)}$, that uses local linear smoothers  of the
periodograms, obtained with least squares, and $d_{W(LK)}$ based on
the exponential transformation of the local  linear smoothers of the
log-periodograms, obtained using maximum likelihood.

The other two dissimilarity measures are based on statistics that
were originally introduced to test the equality of the log spectra
of two processes (see \citet{Vilar2} for details). Let $m_X(\lambda)
= \log(f_X(\lambda))$ and similarly for $Y$. The measures considered
were
$$
d_{GLK}(X,Y) = \sum_{k=1}^n \Big[Z_k-\hat\mu(\lambda_k) -
2\log\big(1+e^{(Z_k-\hat\mu(\lambda_k))}\big)\Big]
 -\sum_{k=1}^n [Z_k -2\log(1+e^{Z_k})],
$$
where $Z_k=\log(I_X(\lambda_k))-\log(I_Y(\lambda_k))$,
$\mu(\lambda_k) = m_X(\lambda_k) -m_Y(\lambda_k)$ and
$\hat\mu(\lambda_k)$ is the local maximum log-likelihood estimator
of $\mu(\lambda_k)$ computed by local linear fitting, and
$$
d_{ISD}(X,Y) = \int \big(\hat m_x(\lambda)-\hat
m_Y(\lambda)\big)^2\, d\lambda,
$$
where $\hat m_X(\lambda)$ and $\hat m_Y(\lambda)$ are the local
linear smoothers of the log-periodograms, obtained using the maximum
local likelihood in this case. It is important to observe

Four series from each spectrum were simulated with the purpose of
testing whether the different criteria were able to recover the
original groups. Table \ref{tabla:Exp3} gives the results.
In this case the method proposed in this work performs
better than the rest for short series, followed closely by ACFG. For
medium sized series the best results correspond to $d_{W(LK)}$ while
for long series $T=1000$ several methods, the TV distance among
them, work equally well. In this experiment, in
general, methods not using logarithms perform better than those that
use it.  It is important to observe that the  methods
that use the likelihood function are very slow for long series.

\renewcommand{\arraystretch}{1}
\begin{table}
\centering \tiny
\begin{tabular}{  l|l l l l l l l l l l l l l}
\multicolumn{14}{l}{$T=100,\ k=2$}\\
\hline \hline
N&  ACFG&   ACFU&   P&  NP& LP& LNP&    CEP&  TV&   L$^1$ & W(DLS) & W(LK) & ISD & GLK\\
\hline
100&    0.783&  0.769&  0.623&  0.764&  0.669&  0.662&  0.654& \underline{0.786}& 0.641 & 0.636 & 0.750 & 0.742 & 0.721\\
500&    0.785&  0.771&  0.624&  0.762&  0.671&  0.655&  0.669& \underline{0.790}& 0.657 & 0.641 & 0.736 & 0.731 & 0.711\\
\hline \multicolumn{14}{l}{\ } \\
\multicolumn{14}{l}{$T=200,\ k=2$}\\
\hline \hline
N&  ACFG&   ACFU&   P&  NP& LP& LNP&    CEP&  TV&   L$^1$ & W(DLS) & W(LK) & ISD & GLK\\
\hline
100&    0.879&  0.873&  0.704&  0.874&  0.681&  0.708&  0.677& 0.900& 0.702 & 0.850 & \underline{0.934} & 0.920 & 0.902\\
500&    0.894&  0.875&  0.709&  0.863&  0.706&  0.710&  0.692& 0.905& 0.722 & 0.844 & \underline{0.935} & 0.910 & 0.899\\
\hline  \multicolumn{14}{l}{\ } \\
\multicolumn{14}{l}{$T=1000,\ k=2$}\\
\hline \hline
N&  ACFG&   ACFU&   P&  NP& LP& LNP&    CEP&  TV&   L$^1$ & W(DLS) & W(LK) & ISD & GLK\\
\hline
100&    0.999&  0.999&  0.972&  0.999&  0.818&  0.813&  0.786& \underline{1.000}& 0.944 & 0.996 & \underline{1.000} & 0.995 & \underline{1.000}\\
500&    0.999&  0.999&  0.974&  0.999&  0.855&  0.858&  0.809& \underline{1.000}& 0.943 & 0.994 & \underline{1.000} & 0.996 & \underline{1.000}\\
\hline
\end{tabular}
\caption{Results for the Experiment 3. $T$ is the length of the
series, $k$ the number of clusters and $N$ the number of
replications. The best result for each value of $T$ is underlined.}
\label{tabla:Exp3}
\end{table}

\section{Applications.}

As was mentioned in the introduction, an interesting and important
problem in Oceanography is the determination of stationary sea
states. Consider a time series that represents the sea surface
height at a fixed point as a function of time.   If the sea state is
stationary, the spectrum of this time series can be interpreted as
the distribution of the energy as a function of frequency for the
given sea state.

Typically, stationary sea states last for some time (hours or days),
and then, due to changing weather conditions, sea currents, the
presence of swell or other reasons, change to a different state.
These changes do not occur instantaneously, but rather require a
certain time to develop, during which there is a transition between
the initial and final states. In this context, the usual
segmentation methods that seek to determine change-points in a
non-stationary time series do not work well, and a different point
of view for the problem may be helpful: instead of looking for
change-points, the idea is to identify short stationary intervals
which have similar behavior, in terms of their spectral densities.
If these intervals are contiguous in time, then it is reasonable to
assume that they constitute a single (longer) stationary interval.
The similarity is determined using the TV distance over normalized
spectra and the clustering procedure described in section
\ref{sec3}.

\subsection{Simulated Data}\label{simdata}

Further simulation studies were carried out to assess the
performance of the clustering algorithm in the presence of
transition periods. The main objective was to gauge the performance
when slow transitions between stationary periods are present in a
data set. The simulations were carried out using the JONSWAP and
Torsethaugen families of spectra. The latter is a family of bimodal
spectra used in Oceanography, which accounts for the presence of
swell and wind-generated waves, and was also developed to model
spectra observed in North-Sea locations. Details can be found in
\citet{Torset1} and \citet{Torset2}.

In all cases the significant wave height ($H_s$) was set to 1; the
simulated series starts with waves from a stationary period of 4
hours (JONSWAP spectrum with peak period $T_p=3.6$), then a
transition lasting 3 hours to another stationary period (JONSWAP
spectrum with $T_p = 4.2$) and, after 4 hours, a new 3-hour
transition to a third 4-hour stationary period (Torsethaugen
spectrum with $T_p = 5.0$). Figure \ref{S_Exp3} (right) shows the
 spectra involved in the experiment. One thousand
replications of this scheme were simulated. The description of the
method for simulating transition periods can be seen in \citet{EOyA1}.

\begin{table}
 \centering
        {
 \centering
  \subfloat[\label{Tab_JJT5C}]
  {
\tiny
\renewcommand{\arraystretch}{1}
\begin{tabular}{|r|r|r|r|r|r|}
\hline
 & \multicolumn{5}{c|}{Cluster}\\
\hline
& 1 & 2 & 3 & 4 & 5\\
\hline \rowcolor{Coral3}
1 & 1000 &  &  &  & \\
\hline \rowcolor{Coral3}
2 & 1000 &  &  &  & \\
\hline \rowcolor{Coral3}
3 & 1000 &  &  &  & \\
\hline \rowcolor{Coral3}
4 & 1000 &  &  &  & \\
\hline \rowcolor{Coral3}
5 & 1000 &  &  &  & \\
\hline \rowcolor{Coral3}
6 & 1000 &  &  &  & \\
\hline \rowcolor{Coral3}
7 & 1000 &  &  &  & \\
\hline \rowcolor{Coral3}
8 & 999 & 1 &  &  & \\
\hline \rowcolor{CadetBlue1}
9 & 981 & 19 &  &  & \\
\hline \rowcolor{CadetBlue1}
10 & 533 & 467 &  &  & \\
\hline \rowcolor{CadetBlue1}
11 & 174 & 826 &  &  & \\
\hline \rowcolor{CadetBlue1}
12 &  & 906 & 88 &  & \\
\hline \rowcolor{CadetBlue1}
13 &  & 487 & 513 &  & \\
\hline \rowcolor{CadetBlue1}
14 &  & 72 & 926 & 2 & \\
\hline \rowcolor{DarkSeaGreen3}
15 &  & 45 & 950 & 5 & \\
\hline \rowcolor{DarkSeaGreen3}
16 &  & 41 & 953 & 6 & \\
\hline \rowcolor{DarkSeaGreen3}
17 &  & 47 & 947 & 6 & \\
\hline \rowcolor{DarkSeaGreen3}
18 &  & 44 & 952 & 4 & \\
\hline \rowcolor{DarkSeaGreen3}
19 &  & 43 & 952 & 5 & \\
\hline \rowcolor{DarkSeaGreen3}
20 &  & 43 & 950 & 7 & \\
\hline \rowcolor{DarkSeaGreen3}
21 &  & 45 & 951 & 4 & \\
\hline \rowcolor{DarkSeaGreen3}
22 &  & 44 & 951 & 5 & \\
\hline \rowcolor{AntiqueWhite1}
23 &  & 29 & 939 & 32 & \\
\hline \rowcolor{AntiqueWhite1}
24 &  &  & 464 & 536 & \\
\hline \rowcolor{AntiqueWhite1}
25 &  &  & 127 & 873 & \\
\hline \rowcolor{AntiqueWhite1}
26 &  &  &  & 824 & 176\\
\hline \rowcolor{AntiqueWhite1}
27 &  &  &  & 449 & 551\\
\hline \rowcolor{AntiqueWhite1}
28 &  &  &  & 21 & 979\\
\hline \rowcolor{DarkOrchid4!50}
29 &  &  &  & 1 & 999\\
\hline \rowcolor{DarkOrchid4!50}
30 &  &  &  &  & 1000\\
\hline \rowcolor{DarkOrchid4!50}
31 &  &  &  & 1 & 999\\
\hline  \rowcolor{DarkOrchid4!50}
32 &  &  &  & 1 & 999\\
\hline  \rowcolor{DarkOrchid4!50}
33 &  &  &  & 1 & 999\\
\hline  \rowcolor{DarkOrchid4!50}
34 &  &  &  & 1 & 999\\
\hline  \rowcolor{DarkOrchid4!50}
35 &  &  &  & 2 & 998\\
\hline  \rowcolor{DarkOrchid4!50}
36 &  &  &  &  & 1000\\
\hline
\end{tabular}
}
 \qquad
  \subfloat[ \label{Tab_JJT3C}]
  {
\tiny
\renewcommand{\arraystretch}{1}
\begin{tabular}{|r|r|r|r|}
\hline
 & \multicolumn{3}{c|}{Cluster}\\
\hline
& 1 & 2 & 3\\
\hline \rowcolor{Coral3}
1 & 1000 &  & \\
\hline \rowcolor{Coral3}
2 & 1000 &  & \\
\hline \rowcolor{Coral3}
3 & 1000 &  & \\
\hline \rowcolor{Coral3}
4 & 1000 &  & \\
\hline \rowcolor{Coral3}
5 & 1000 &  & \\
\hline \rowcolor{Coral3}
6 & 1000 &  & \\
\hline \rowcolor{Coral3}
7 & 1000 &  & \\
\hline \rowcolor{Coral3}
8 & 1000 &  & \\
\hline \rowcolor{CadetBlue1}
9 & 1000 &  & \\
\hline \rowcolor{CadetBlue1}
10 & 980 & 20 & \\
\hline \rowcolor{CadetBlue1}
11 & 815 & 185 & \\
\hline \rowcolor{CadetBlue1}
12 & 565 & 435 & \\
\hline \rowcolor{CadetBlue1}
13 & 188 & 812 & \\
\hline \rowcolor{CadetBlue1}
14 & 5 & 995 & \\
\hline \rowcolor{DarkSeaGreen3}
15 & 1 & 999 & \\
\hline \rowcolor{DarkSeaGreen3}
16 &  & 1000 & \\
\hline \rowcolor{DarkSeaGreen3}
17 &  & 1000 & \\
\hline \rowcolor{DarkSeaGreen3}
18 &  & 1000 & \\
\hline \rowcolor{DarkSeaGreen3}
19 &  & 1000 & \\
\hline \rowcolor{DarkSeaGreen3}
20 &  & 1000 & \\
\hline \rowcolor{DarkSeaGreen3}
21 &  & 1000 & \\
\hline \rowcolor{DarkSeaGreen3}
22 &  & 1000 & \\
\hline \rowcolor{AntiqueWhite1}
23 &  & 1000 & \\
\hline \rowcolor{AntiqueWhite1}
24 &  & 863 & 137\\
\hline \rowcolor{AntiqueWhite1}
25 &  & 585 & 415\\
\hline \rowcolor{AntiqueWhite1}
26 &  & 297 & 703\\
\hline \rowcolor{AntiqueWhite1}
27 &  & 57 & 943\\
\hline \rowcolor{AntiqueWhite1}
28 &  &  & 1000\\
\hline \rowcolor{DarkOrchid4!50}
29 &  &  & 1000\\
\hline \rowcolor{DarkOrchid4!50}
30 &  &  & 1000\\
\hline \rowcolor{DarkOrchid4!50}
31 &  &  & 1000\\
\hline \rowcolor{DarkOrchid4!50}
32 &  &  & 1000\\
\hline \rowcolor{DarkOrchid4!50}
33 &  &  & 1000\\
\hline \rowcolor{DarkOrchid4!50}
34 &  &  & 1000\\
\hline \rowcolor{DarkOrchid4!50}
35 &  &  & 1000\\
\hline \rowcolor{DarkOrchid4!50}
36 &  &  & 1000\\
\hline
\end{tabular}
}
 \qquad
\subfloat[\label{Tab_JJT5A}]
  {
\tiny
\renewcommand{\arraystretch}{1}
\begin{tabular}{|r|r|r|r|r|r|}
\hline
 & \multicolumn{5}{c|}{Cluster}\\
\hline
& 1 & 2 & 3 & 4 & 5\\
\hline \rowcolor{Coral3}
1 & 1000 &  &  &  & \\
\hline \rowcolor{Coral3}
2 & 1000 &  &  &  & \\
\hline \rowcolor{Coral3}
3 & 1000 &  &  &  & \\
\hline \rowcolor{Coral3}
4 & 1000 &  &  &  & \\
\hline \rowcolor{Coral3}
5 & 1000 &  &  &  & \\
\hline \rowcolor{Coral3}
6 & 1000 &  &  &  & \\
\hline \rowcolor{Coral3}
7 & 1000 &  &  &  & \\
\hline \rowcolor{Coral3}
8 & 1000 &  &  &  & \\
\hline \rowcolor{CadetBlue1}
9 & 991 & 1 &  &  & \\
\hline \rowcolor{CadetBlue1}
10 & 559 & 441 &  &  & \\
\hline \rowcolor{CadetBlue1}
11 & 168 & 832 &  &  & \\
\hline \rowcolor{CadetBlue1}
12 &  & 872 & 126 &  & \\
\hline \rowcolor{CadetBlue1}
13 &  & 421 & 579 &  & \\
\hline \rowcolor{CadetBlue1}
14 &  & 68 & 931 & 1 & \\
\hline \rowcolor{DarkSeaGreen3}
15 &  & 49 & 950 & 1 & \\
\hline \rowcolor{DarkSeaGreen3}
16 &  & 49 & 950 & 1 & \\
\hline \rowcolor{DarkSeaGreen3}
17 &  & 49 & 950 & 1 & \\
\hline \rowcolor{DarkSeaGreen3}
18 &  & 49 & 950 & 1 & \\
\hline \rowcolor{DarkSeaGreen3}
19 &  & 49 & 950 & 1 & \\
\hline \rowcolor{DarkSeaGreen3}
20 &  & 49 & 950 & 1 & \\
\hline \rowcolor{DarkSeaGreen3}
21 &  & 49 & 950 & 1 & \\
\hline \rowcolor{DarkSeaGreen3}
22 &  & 49 & 950 & 1 & \\
\hline \rowcolor{AntiqueWhite1}
23 &  & 41 & 945 & 14 & \\
\hline \rowcolor{AntiqueWhite1}
24 &  &  & 498 & 502 & \\
\hline \rowcolor{AntiqueWhite1}
25 &  &  & 126 & 873 & 1\\
\hline \rowcolor{AntiqueWhite1}
26 &  &  &  & 815 & 185\\
\hline \rowcolor{AntiqueWhite1}
27 &  &  &  & 415 & 585\\
\hline \rowcolor{AntiqueWhite1}
28 &  &  &  & 5 & 995\\
\hline \rowcolor{DarkOrchid4!50}
29 &  &  &  &  & 1000\\
\hline \rowcolor{DarkOrchid4!50}
30 &  &  &  &  & 1000\\
\hline \rowcolor{DarkOrchid4!50}
31 &  &  &  &  & 1000\\
\hline  \rowcolor{DarkOrchid4!50}
32 &  &  &  &  & 1000\\
\hline  \rowcolor{DarkOrchid4!50}
33 &  &  &  &  & 1000\\
\hline  \rowcolor{DarkOrchid4!50}
34 &  &  &  &  & 1000\\
\hline  \rowcolor{DarkOrchid4!50}
35 &  &  &  & 1 & 999\\
\hline  \rowcolor{DarkOrchid4!50}
36 &  &  &  &  & 1000\\
\hline
\end{tabular}
}
 \caption{Results for the simulated transition data using:
  a) Complete linkage function and $5$ groups, b)
  Complete linkage function and $3$ groups,
  c) Average linkage function and $5$ groups.}\label{tab:chapter4:1}
}
\end{table}

The simulated data was divided into 30-minute intervals and the
corresponding spectra were estimated. Using these spectral
densities, the clustering algorithm based on the TV distance
 was applied with two different linkage
functions, complete and average. The  algorithm was expected to
recover five groups, the three stationary periods and two
transitions. Table \ref{tab:chapter4:1} shows the results for the
1000 replications. The row color stands for the original groups, for
example, intervals $1$ to $8$, colored in dark orange, represent the
first group. The column corresponds to the group assigned by the
clustering algorithm.

Table  \ref{tab:chapter4:1}(a) shows the results with the complete
link function and  5 clusters. It can be seen that for the initial
and final stationary periods, the algorithm almost always gives the
correct result. For the central stationary period the success rate
is around 95\%. The algorithm has a harder time identifying the
transition periods, which is reasonable since these are not
homogeneous groups. In particular, intervals at the beginning and
end of a transition period are classified as belonging to the
nearest stationary period over 90\% of the time in all cases. This
is also reasonable since the transition is slow and due to the
sampling variability in the estimation of the spectral densities,
such small differences are difficult to detect.   Table
\ref{tab:chapter4:1}(c) shows the results for the average link
function, which are similar.

One could argue that, in fact, the transition periods should not be
considered as separate clusters, since they do not correspond to
time intervals having homogeneous spectral densities, and in
consequence one should only consider three groups. Table
\ref{tab:chapter4:1}(b) shows the results in this case for the
complete link function. Almost always, the stationary groups are
correctly assigned to the same group.  Transition intervals tend to
be classified in the closest stationary group.  These results could
be used in a two-tier process, in which, in a given realization and
using the results of the clustering algorithm, the intervals at the
border would be tested to decide whether they really belong to the
same group as the rest, or they should be considered as belonging to
a transition period and moved outside the cluster. This idea will
not be further developed in this work.

The initial division of the time series into 30-minute intervals
determines the precision with which the stationary intervals can be
determined. Shorter intervals will increase it but, on the other
hand, using less data to estimate the spectral density will increase
the statistical variability of the estimation. To test whether
shorter time intervals would give good results, the simulated series
were analyzed dividing them into 20-minute intervals. In general,
when dealing with unimodal spectral densities such as those from the
JONSWAP family, results were similar, but for bimodal densities in
the Torsethaugen family results were worse. In this case the
procedure has difficulty in clustering together the estimated
spectra from this family, resulting in an important decrease in the
number of correct classifications. For this reason, intervals of 30
minutes were used for the analysis of real data in the next section.

\subsection{Real Data Analysis}\label{sec5-2}

Our starting point in this section is the idea that sea states at a
fixed point on the sea surface can be modeled as a sequence of
alternating stationary and transition periods. With this structure
in mind and based on  the results of the simulations shown in
Section \ref{simdata}, we carried out a clustering analysis over a
real data set in order to detect these periods.

We used real wave data obtained from the U. S. Coastal Data
Information Program (CDIP) website. The data come from buoy 106
(51201 for the National Data Buoy Center), located in Waimea Bay,
Hawaii, at a water depth of 200 m. and correspond to 192 30-minute
intervals starting on January 1st., 2003, a total of 96 hours (4
days).

\begin{figure}[t]
\centering
 \includegraphics[scale=0.4]{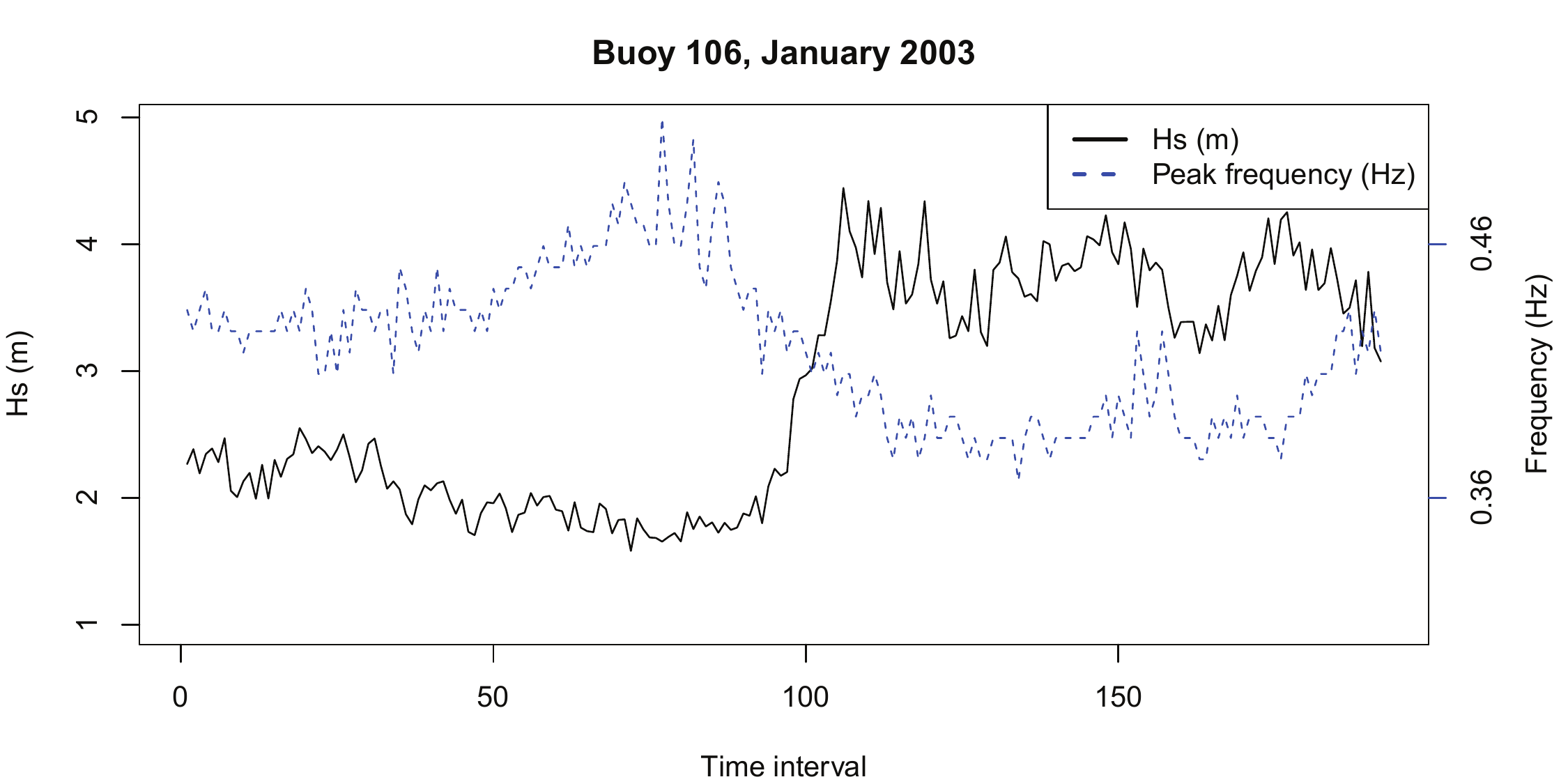}
 \caption{Significant wave height (solid) and peak frequency (dashed) for buoy 106, Jan 1st to 4th, 2003.}
 \label{Eng_Buoy106Jan03_V2_Hs_1_192}
\end{figure}

Figure \ref{Eng_Buoy106Jan03_V2_Hs_1_192} shows both significant
wave height (solid) and spectral peak frequency (dashed) for this
data set. It shows that $H_s$ starts with values around 2 m. and
then, about the middle of the time interval, increases in a few
hours to values around 3.7-4 m. where it remains for the rest of the
period. On the other hand, the spectral peak frequency starts the
period slightly increasing, then starts to decrease as $H_s$
increases, to remain low for the rest of the period.

The clustering analysis was carried out in two different ways.
Initially the complete data set, comprising the 192 time intervals,
was considered. Alternatively the data set was divided into two
groups, group 1 including intervals 1 - 86 and group 2 intervals 87
- 192. A comparison of the results obtained in each case gives
indications about the consistency of the proposed method and also
allows for the evaluation of possible boundary effects in the
segmentation procedure.

As in Section \ref{simdata}, for each 30-minute interval the
spectral density was estimated and normalized, and the matrix of
total variation distances between these spectra was calculated. This
matrix is the input for the agglomerative hierarchical clustering
procedure. We tried two of the main linkage functions: complete and
average, with similar results.

Unlike the simulations of Section \ref{simdata} where the number of
clusters is known, now this number is unknown. In order to decide
the appropriate number of clusters, $k$, to be considered in the
analysis a suitable method should be chosen. A good review of
available methods to assess the value of $k$ can be found, for
example, in chapter 17 of \citet{Gan}. In our case, as we have no
external information about the number and definition of clusters we
must select an internal method. Two methods were used, Dunn and
Davies-Bouldin,  with very similar results. We present here the
results obtained with Dunn's index, which is defined as
$$
V_D(k)= \min_{1\le i \le k} \left\lbrace \min_{i+1\le j \le k}
\left( \frac{D(C_i,C_j)}{\max_{1\le h \le k} diam(C_h)} \right)
\right\rbrace,
$$
where $k$ is the number of clusters, $D(C_i,C_j)$ is the distance
between clusters $C_i$ and $C_j$, and $diam(C_h)$ is the diameter of
cluster $C_h$.

\begin{figure}[t]
\centering
 \includegraphics[scale=0.65]{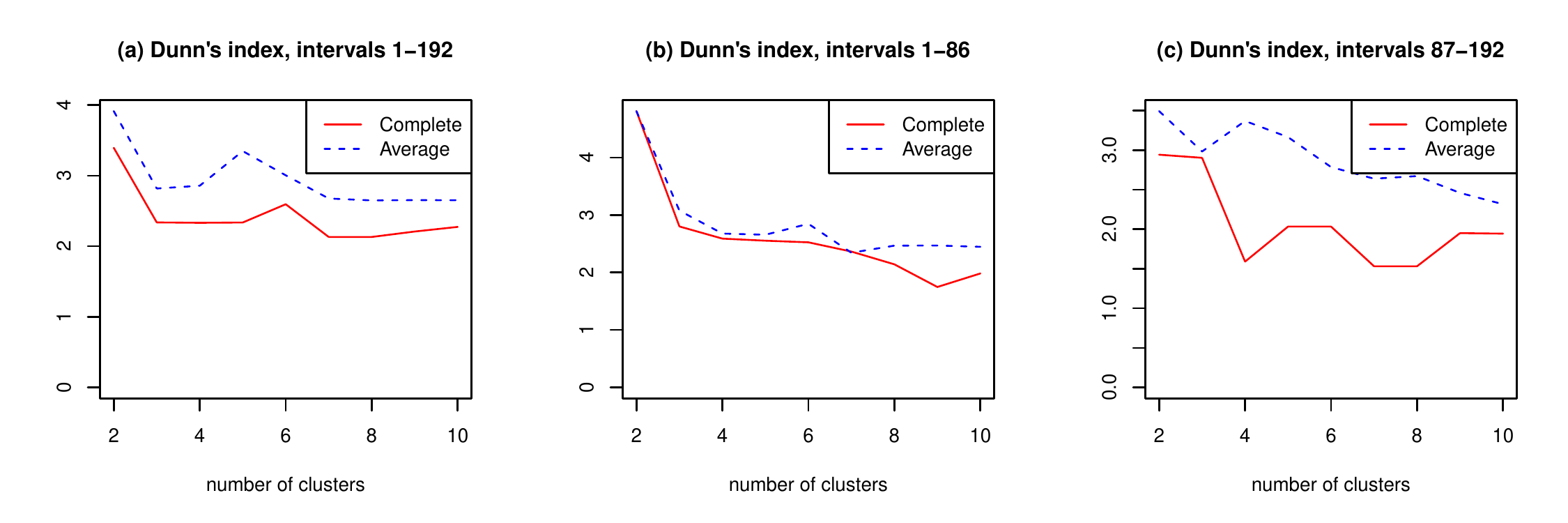}
 \caption{Dunn's index for the three clustering processes using complete and average linkage functions:
 (a, left) intervals 1-192, (b, middle) intervals 1-86 and (c, right) intervals 87-192.}
 \label{Eng_Dunn_los3}
\end{figure}

From the definition of $V_D$ it is clear that high values point to
suitable values of $k$. The computation of this index was carried
out using the \textit{clv} package in R \citep{R}. Among the
available metrics to compute $D(C_i,C_j)$ and $diam(C_h)$ the
average was selected. The results for the three clustering
procedures and the two linkage functions are shown in Figure
\ref{Eng_Dunn_los3}. The main conclusion is that there is a good
degree of agreement between the two linkage functions for each
clustering procedure. Plot (a) for the clustering over the whole 192
 intervals indicates the existence of 5 (average) or 6
(complete) groups. Dunn's index in plot (b) for the first 86-time
periods indicates clearly the existence of two groups, and finally,
plot (c) for the second half of the period points to 3 (complete) or
4 (average) groups, depending on the linkage function.

\begin{figure}[t]
\centering
 \includegraphics[scale=0.9]{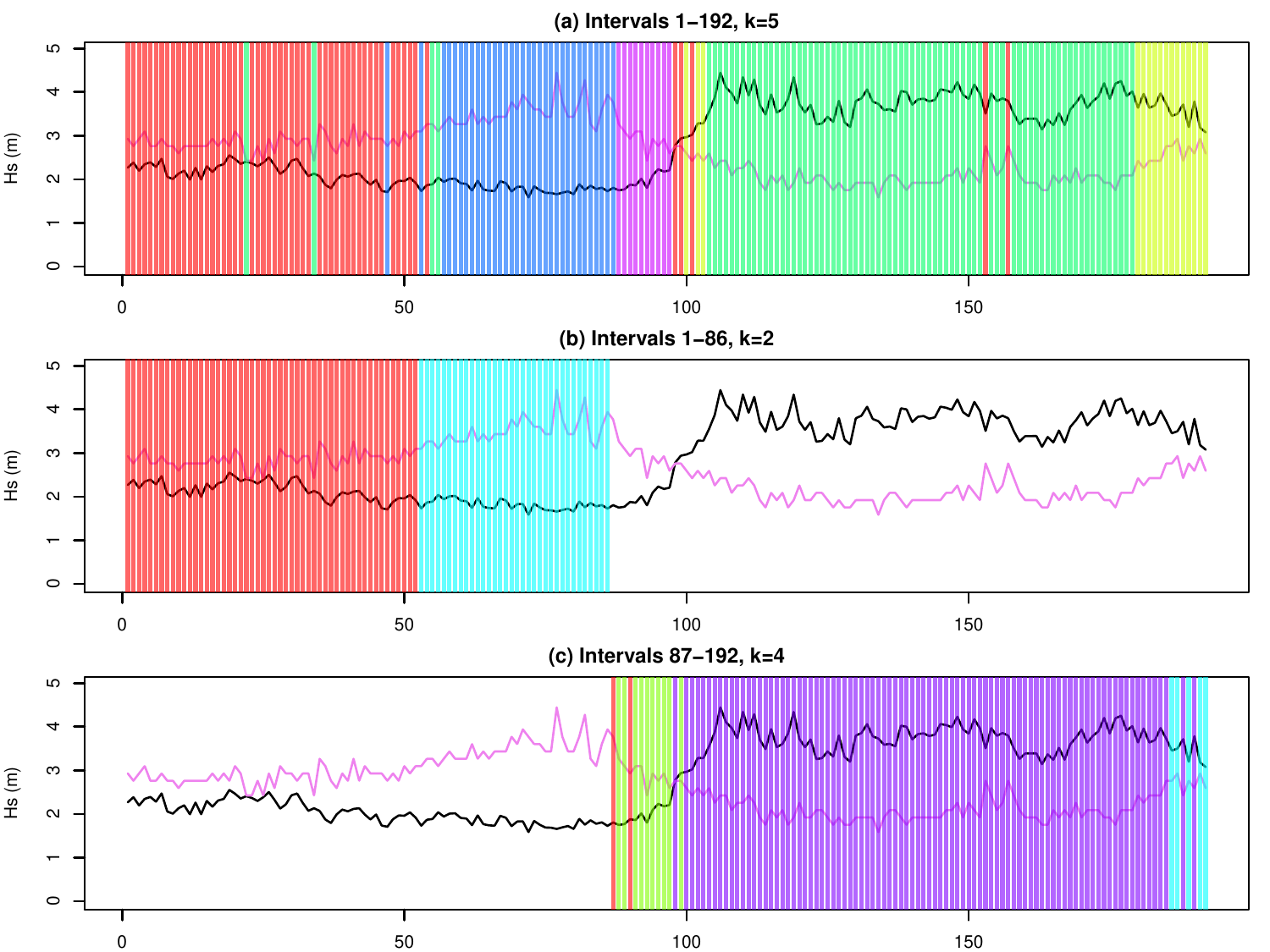}
 \caption{Clustering results using the average linkage function.}
 \label{Eng_Buoy106Jan03_V2_average_clus}
\end{figure}

\begin{figure}[t]
\centering
 \includegraphics[scale=0.9]{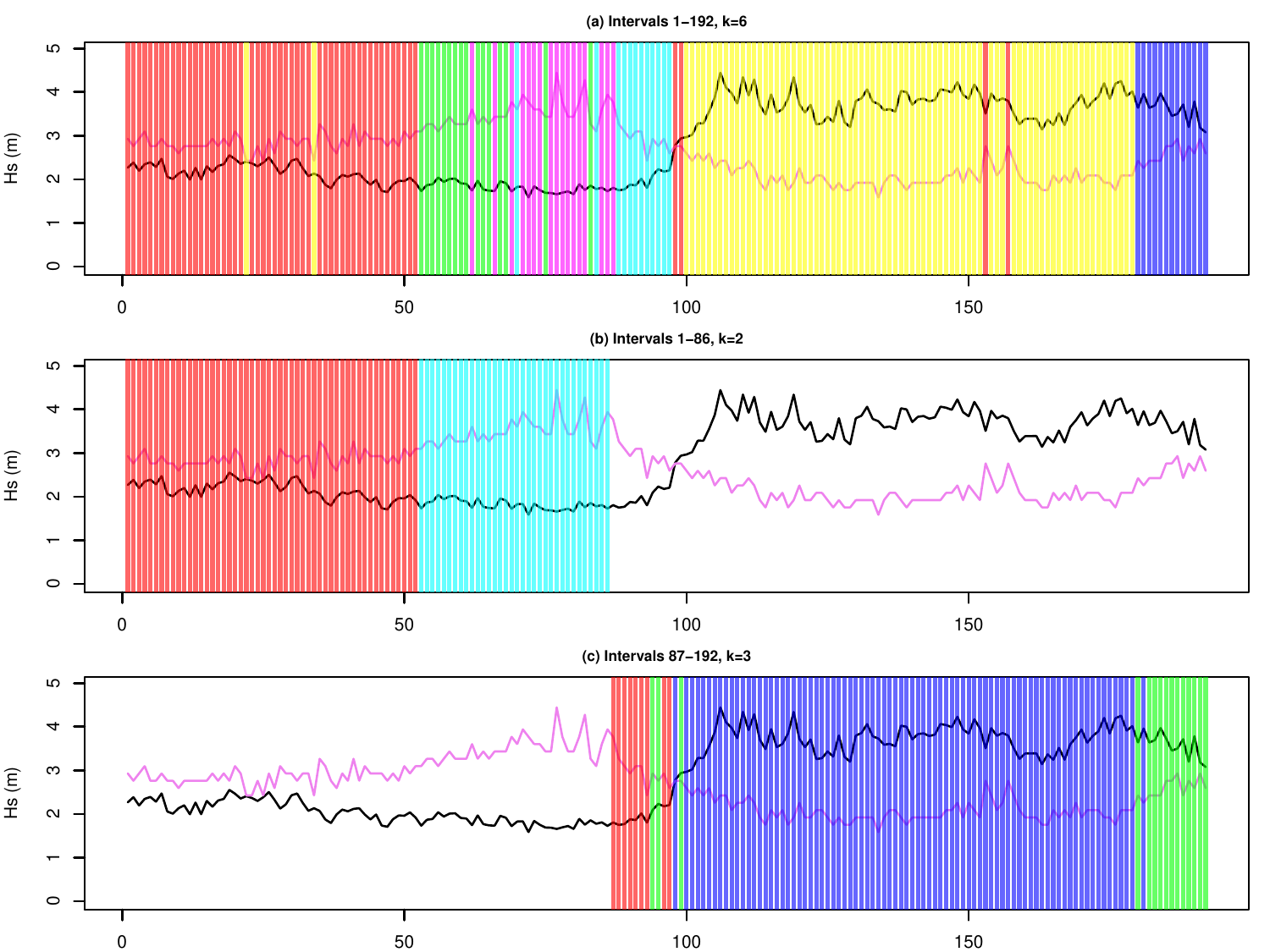}
 \caption{Clustering results using the complete linkage function.}
 \label{Eng_Buoy106Jan03_V2_ward_clus}
\end{figure}

After applying the hierarchical clustering algorithm, a
silhouette analysis was performed to check for possible errors in
the clustering process. In a hierarchical clustering algorithm, once
an element has been assigned to a cluster, it cannot be reassigned
to a different cluster, even if changes in the composition of the
clusters as new elements are incorporated imply that it would have
been better to assign this element to a different group.

The silhouette index, proposed by \citet{Rouss}, gives a measure of
the adequacy of each point to its cluster. Let $a(i)$ be the average
distance or dissimilarity of point $i$ with all the other elements
within the same cluster, and let $b(i)$ be the smallest average
dissimilarity of $i$ to any of the clusters to which $i$ does not
belong. Then the silhouette index of $i$ is defined as
$$
s(i) = \frac{b(i)-a(i)}{\max\{a(i), b(i)\}}
$$
This index satisfies $-1 \leq s(i) \leq 1$ for all $i$, and large
positive values indicate that the element has been well classified
while negative values point to misclassification. As a consequence
the classification of intervals with negative silhouette index was
revised. The reclassification only affected between 6 and 9
intervals, for the different cases considered.

Figures \ref{Eng_Buoy106Jan03_V2_average_clus} and
\ref{Eng_Buoy106Jan03_V2_ward_clus} show the results of the
clustering procedure for the average and complete linkage functions,
respectively, and $k$ chosen using Dunn's index, after revision with
the silhouette index. The first
 point to note in both figures is that, broadly speaking,
the clustering procedure captures the time structure in the data. In
other words, using only information about the TV distance between
normalized spectral densities, the clustering procedure groups in
the same cluster intervals that are contiguous in time, and this is
valid except for a few intervals in each case.

 Plot (a) of Figure \ref{Eng_Buoy106Jan03_V2_average_clus}
shows the groups for the whole time interval using $k=5$ clusters
and the average linkage function. In this plot, the period of time
between intervals 1 and 87, just before $H_s$ starts to increase, is
essentially divided into two groups. The first cluster (red)
comprises most of the initial 54 time intervals, with the exception
of four of them, 22 and 34, which belong to the fourth cluster and
47 and 53, which belong to the second. It corresponds to a period of
time during which both $H_s$ and $T_p$ are stable. The second
cluster (57-87, in blue) includes the rest of the intervals in the
first half and represents an uninterrupted sequence of time
intervals, except for intervals 55 and 56, which are assigned to the
fourth cluster. This is very similar to the clustering obtained for
time intervals 1 - 86, represented in plot (b). In this case there
are only two groups and the main difference with the first half of
plot (a) is the starting point of the second cluster, which has
moved to the left, to interval 53. The other difference is that now
intervals 22, 34 and 47 are assigned to the first cluster, which
becomes a single block, while the rest forms a second
block.

The third group (88-97, purple) in plot (a) corresponds to the
initial stages of growth for $H_s$, and is followed by 6 intervals,
three of which (98, 99 and 101) belong to cluster 1 while the rest
belong to cluster 5. The fourth cluster (104-179, green except 153
and 157) is the largest and includes almost all intervals of the
period where $H_s$ oscillates around 3.7-4 m. There are two red
intervals that break the time continuity of this cluster, 153 and
157, which belong to cluster 1. Finally, the fifth cluster (180-192,
yellow) appears at the end of this last period, and is also an
uninterrupted sequence of time intervals. Comparing with plot (c),
which corresponds to intervals 87-192 divided into 4 groups, we see
that the first interval (red) is classified as a cluster together
with interval 90. In plot (a) interval 87 is the last in cluster 2
(dark blue). The second cluster (green) in plot (c) has similarities
with cluster 3 in plot (a). The third cluster (purple) encompasses
the fourth cluster (green) in plot (a) plus the segments included in
cluster 1 (red) for this half of the data, as well as 8 intervals
included in cluster 5 (yellow). Finally, cluster 4 (light blue) in
plot (b) groups the rest of the intervals in cluster 5 (yellow) of
plot (a).

As can be seen from this analysis, although there are some
differences between the clustering obtained for the whole data set
and those of the two halves, in general the agreement is very good,
and those intervals in which the clustering differs, probably
correspond either to transition intervals, such as the final
intervals in plots (a) and (c), or to intervals in which temporary
changes in the sea conditions (the presence of swell or local
variations in the wind, for example) produce changes that disappear
once these temporary conditions cease, as may be the case for
intervals  22, 34, 153 and 157.  A possible conclusion
from this analysis is that there are three stable periods: 1 - 52,
57 - 87 and 104 - 179, and the other intervals correspond to
transition periods.

A similar analysis can be done for the three plots in
Figure \ref{Eng_Buoy106Jan03_V2_ward_clus}, which correspond to the
results with the complete linkage function. Plot (a) shows the
complete time interval with 6 groups, while plots (b) and (c)
correspond to the two halves with 2 and 3 clusters, respectively, as
suggested by Dunn's index.

For the first half, the first group in plots (a) and (b) is almost
the same, except for intervals 22 and 34, which are assigned to a
different group in plot (a). The assignement of intervals 52-86,
however, is quite different. While in plot (b) they are all in a
single cluster, in plot (a) they are divided into 3 different
groups, none of which appears as a single block in time. As regards
the second half, again the largest group (100-179) is very similar
in both graphs, the difference being two intervals (153 and 157) in
plot (a) that are assigned to a different group. The remaining
intervals, 87-99 and 180-192 also show very similar structures in
both cases.

Comparing now Figures 6 (a) and 7 (a) we see that the main
differences lie in the interval 53-103, while the rest match very
well. These differences are partly due to the fact that 5 clusters
were chosen for the average linkage function, while for the complete
the choice was 6 clusters. We also tried the complete linkage
function with 5 clusters (not shown) and this shows more
similarities with figure \ref{Eng_Buoy106Jan03_V2_average_clus}(a).
On the other hand, the average linkage function seems to produce
clusterings that are more homogeneous in time than those obtained
using the complete link, although further research in this respect
is needed.

\section{Conclusions}

In this paper, a new method for time series clustering was proposed.
The method is based on using the total variation distance between
normalized spectra as a measure of dissimilarity between time
series. Simulation results (Sec. 4) show that the method has a
performance that is comparable to the best clustering methods based
on features extracted from the raw data, and in certain cases it
performs better than the rest. Simulations (Sec. \ref{simdata}) also
show that the method is capable of detecting stationary periods in
situations where slow transitions between stationary states occur.

The method was used for the analysis of real sea wave data, measured
at a fixed location, with the purpose of detecting stationary and
transition periods. The results obtained using the average and
complete linkage functions, presented in Sec. 5.2, show a reasonable
agreement for the two linkage functions, taking into account that
the number of clusters suggested by Dunn's index is different. The
analysis also shows that the results are consistent when the
clustering method is applied over intervals of different length.

However, further research is needed to find a better criterion for
choosing the number of clusters. Other aspects that need a closer
look are the optimal length of the time window and the overlap
interval for the automatic segmentation of longer data series, the
use of trimming in order to robustify the clustering process and the
possibility of using functional clustering with the spectral
densities, instead of using the TV distance.

\section{Acknowledgements}

The software WAFO developed by the Wafo group at Lund University of
Technology, Sweden, available at
http://www.maths.lth.se/matstat/wafo was used for the calculation of
all spectral densities and associated spectral characteristics. The
data for station 106 were furnished by the Coastal Data Information
Program (CDIP), Integrative Oceanographic Division, operated by the
Scripps Institution of Oceanography, under the sponsorship of the
U.S. Army Corps of Engineers and the California Department of
Boating and Waterways (http://cdip.ucsd.edu/).

 This work was partially supported by CONACYT, Mexico, Proyecto
 An\'alisis Estad\'{\i}stico de Olas Marinas, Fase II. J. Ortega
 wishes to thank Prof. Adolfo J. Quiroz for several fruitful
 conversations on the topic of this paper. P.C. Alvarez Esteban
 wishes to acknowledge CIMAT, A.C., the Spanish Ministerio de Ciencia
 y Tecnolog\'{\i}a, grants MTM2011-28657-C02-01 and
 MTM2011-28657-C02-02 and the Consejer\'{\i}a de Educaci\'on de la
 Junta de Castilla y Le\'on, grant VA212U13 for their financial
 support.

\end{document}